\documentclass[iop,revtex4]{emulateapj-rtx4}
\usepackage{epsfig}
\usepackage{amsmath}
\usepackage{natbib, aas_macros}
\usepackage{lscape}
\begin{document}

\title{NIR spectroscopic observation of massive galaxies in the protocluster at $z=3.09$}
\author{Mariko Kubo\altaffilmark{1, 2}, Toru Yamada\altaffilmark{2}, Takashi Ichikawa\altaffilmark{2}, Masaru Kajisawa\altaffilmark{3}, Yuichi Matsuda\altaffilmark{4, 5}, Ichi Tanaka\altaffilmark{6} }
%
%
\altaffiltext{1}{ Institute for Cosmic Ray Research, University of Tokyo, 5-1-5 Kashiwa-no-Ha, Kashiwa City
Chiba, 277-8582, Japan}
\altaffiltext{2}{Astronomical Institute, Tohoku University, 6-3 Aoba, Aramaki, Aoba-ku, Sendai, Miyagi, 980-8578, Japan}
\altaffiltext{3}{Research Center for Space and Cosmic Evolution, Ehime University, Bunkyo-cho 2-5, Matsuyama 790-8577, Japan}
\altaffiltext{4}{Chile Observatory, National Astronomical Observatory of Japan, Tokyo 181-8588, Japan}
\altaffiltext{5}{Graduate University for Advanced Studies (SOKENDAI), Osawa 2-21-1, Mitaka, Tokyo 181-8588, Japan}
\altaffiltext{6}{ Subaru Telescope, National Astronomical Observatory of Japan, 650 North A’ohoku Place, Hilo, HI 96720, USA }

\begin{abstract} 
We present the results of near-infrared spectroscopic observations  
of the $K$-band selected candidate galaxies  
in the protocluster at $z=3.09$ in the SSA22 field. 
We observed 67 candidates with $K_{\rm AB}<24$ 
and confirmed redshifts of the 39 galaxies at $2.0< z_{\rm spec}< 3.4$. 
Of the 67 candidates, 24 are certainly protocluster members with $3.04\leq z_{\rm spec}\leq 3.12$,  
which are massive red galaxies those have been unidentified 
in previous optical observations of the SSA22 protocluster. 
Many distant red galaxies (DRGs; $J-K_{\rm AB}>1.4$), 
hyper extremely red objects (HEROs; $J-K_{\rm AB}>2.1$), 
{\it Spitzer} MIPS 24 $\mu$m sources, active galactic nuclei (AGNs)
as well as the counterparts of Ly$\alpha$ blobs and the AzTEC/ASTE 1.1-mm sources 
in the SSA22 field are also found to be the protocluster members. 
The mass of the SSA22 protocluster is estimated  
to be $\sim2-5\times10^{14}~M_{\odot}$ 
and this system is plausibly a progenitor 
of the most massive clusters of galaxies in the current Universe. 
The reddest ($J-K_{\rm AB}\geq 2.4$) protocluster galaxies 
are massive galaxies with $M_{\rm star}\sim10^{11}~M_{\odot}$
showing quiescent star formation activities 
and plausibly dominated by old stellar populations. 
Most of these massive quiescent galaxies host moderately luminous AGNs detected by X-ray. 
There are no significant differences in the 
[O{\footnotesize III}] $\lambda$5007/H$\beta$ emission line ratios, 
and [O{\footnotesize III}] $\lambda$5007 line widths and spatial extents
of the protocluster galaxies from those of massive galaxies at $z\sim2-3$ in the general field.
\end{abstract}


\section{Introduction}
Clusters of galaxies are good evolutionary probes 
for determining the structure formation history in the Universe. 
Number density of massive clusters at given redshift 
is an important quantity for constraining the structure formation model 
as well as the cosmological parameters. 
It is reported that dynamically relaxed, mature clusters 
already existed at $z\sim1.5-2$ (e.g., \citealt{2005ApJ...623L..85M, 2011A&A...526A.133G, 2011A&A...527L..10F}). 
At $z>2$, a number of protoclusters were discovered 
(e.g., \citealt{1998ApJ...492..428S, 2005A&A...431..793V, 
2005ApJ...620L...1O, 2011MNRAS.416.2041M, 2012ApJ...750..137T}) 
although their structures, dynamical properties 
and masses have yet to been revealed well.  
Masses of protoclusters can be measured 
based on e.g., their velocity dispersions and overdensities, 
(e.g., \citealt{2005A&A...431..793V})
but require spectroscopic redshift constraints 
enough to disclose their galaxy distributions.  

On the other hand, protoclusters are suitable targets for studying 
the formation history of massive early-type galaxies 
those dominate cores of massive clusters in the current Universe. 
There is a well-established color-magnitude relation  
among massive early-type galaxies in local clusters
(e.g., \citealt{1977ApJ...216..214V, 1992MNRAS.254..601B}),  
and also in clusters at out to $z\sim1$ 
(e.g., \citealt{1997ApJ...483..582E, 1998ApJ...492..461S}). 
The tight color-magnitude relation implies that 
the bulk of the stars in massive early-type galaxies formed at $z>2$. 
Density excesses of massive red galaxies in protoclusters at $z=2-3$ 
(e.g., \citealt{2007MNRAS.377.1717K, 2012ApJ...750..116U}) 
also suggest that massive early-type galaxies began to assemble 
at as early as $z=2-3$. 
In the meanwhile, massive quiescent galaxies at out to $z\sim5$
were found from field surveys based on color selections  
(e.g., \citealt{2012ApJ...750L..20C};  \citealt{2013A&A...556A..55I};  
\citealt{2013ApJ...777...18M};  \citealt{2014ApJ...783L..14S})
and also confirmed at out to $z\sim3$
\citep{2012ApJ...759L..44G, 2014arXiv1406.0002M}. 
But the relatiohship between such high-z analogues and 
massive early-type galaxies in local clusters remains unclear. 

The protocluster at $z=3.09$ in the SSA22 field is known as 
one of the most outstanding structures at high redshift. 
This protocluster was characterized by the significant over-densities of  
Lyman-break galaxies (LBGs) and Ly$\alpha$ emitters (LAEs)
at $z\approx3.09$ associated with the superstructure extended over $\sim$ 100 Mpc
\citep{1998ApJ...492..428S, 2000ApJ...532..170S, 2004AJ....128.2073H, 2012AJ....143...79Y}. 
The density excesses of Ly$\alpha$ blobs (LABs; \citealt{2004AJ....128..569M}), 
the AzTEC/ASTE 1.1-mm sources \citep{2009Natur.459...61T, 2014MNRAS.440.3462U}
and active galactic nuclei (AGNs) \citep{2009MNRAS.400..299L},  
which may associate with the massive galaxy formation 
were also reported. 

We have studied massive red galaxies in the SSA22 protocluster 
by using our own deep and wide near-infrared (NIR) imaging data taken 
with Multi-Objects Infra-Red Camera and Spectrograph (MOIRCS) 
equipped with Subaru Telescope 
($K\approx 24$ at 5$\sigma$  for 111.8 arcmin$^2$, \citealt{2008PASJ...60..683U, 2012ApJ...750..116U, 2013ApJ...778..170K}). 
We found the surface number density excesses 
of distant red galaxies (DRGs, $J-K>1.4$; \citealt{2003ApJ...587L..79F}), 
hyper extremely red objects (HEROs, $J-K>2.1$; \citealt{2001ApJ...558L..87T})
as well as galaxies with photometric redshift $2.6<z_{\rm phot}<3.6$, 
and those detected with {\it Chandra} and {\it Spitzer} MIPS 24 $\mu$m 
suggesting enhanced AGN and star formation activities. 
On the other hand, up to 50\% of 
the galaxies with stellar mass $M_{\rm star}>10^{11}~M_{\odot}$ at  $2.6<z_{\rm phot}<3.6$ in this field 
have spectral energy distributions (SEDs) dominated by old stellar populations. 

Here we present the results of NIR spectroscopy for the 
$K$-band selected ($K$-selected) candidates of the galaxies in the SSA22 protocluster. 
Our objectives are to confirm the overdensity 
of the $K$-selected galaxies in the SSA22 protocluster, 
investigate their nature and estimate the mass 
of the SSA22 protocluster itself to disclose the 
formation and evolutional history of massive clusters 
and massive galaxies inside clusters. 
The $K$-selected galaxies are massive galaxies 
which should be hosted in massive halos. 
Therefore, they are excellent tracers of cluster mass and 
also very plausible progenitors of massive early-type galaxies. 
Since our targets are faint in rest-frame UV, 
nebular emission lines at rest-frame optical 
are the most useful probes to confirm their redshifts. 
In this paper, we assume cosmological parameters 
of $H_0 = 70$ km s$^{-1}$ Mpc$^{-1}$ , $\Omega_{\rm m} =0.3$ 
and $\Omega_{\Lambda} = 0.7$.  
We use the AB magnitude system throughout this paper. 

\section{Observations and Data reductions}

\begin{table*}[htbp]
 \caption{Summary of the observation}
 \begin{center}
 \begin{minipage}{160mm}
\begin{tabular}{lccccccccc}
\hline \hline
mask ID & R.A. & Dec  & P.A.\footnotemark[a] & UT date & exposure & seeing\footnotemark[b]  & channel & grism & target  \\ 
 & (J2000.0) & (J2000.0) & (degree) & & (sec) & (arcsec) & & & (number)  \\
\hline
SSA22B & 22 17 31.7 & 00 12 40.7 & 222 & 2012 Sep 29 & 14000 & 0.6  & Ch1 & VPH-$K$ & ...\footnotemark[c] \\
 & &  &  &  &  & & Ch2 & VPH-$K$ & 12  \\
SSA22A1 & 22 17 22.6 & 00 17 55.1 & 270 & 2012 Sep 30 & 16000 & 0.7  & Ch1 & VPH-$K$ & ...\footnotemark[c] \\
 &  &  &  &  &  &   & Ch2 & VPH-$K$ & 12 \\
SSA22D & 22 17 26.8 & 00 18 19.4 & 280 & 2012 Oct 27 & 13000 & 0.4  & Ch1 & $HK500$ & 13 \\
 &  &  &  & &  &   & Ch2 & VPH-$K$ & 10 \\
SSA22C & 22 17 30.5 & 00 17 39.8 & 148 & 2012 Oct 28 & 13600 & 0.5  & Ch1 & $HK500$ & 11 \\
 &  &  &  & &  &  & Ch2 & VPH-$K$ & 13  \\
\hline
\footnotetext[a]{ Directions of the slits, from north to east }
\footnotetext[b]{ Average PSF sizes in the $K_s$-band during the observations }
\footnotetext[c]{ Ch1 did not work during the observations on September 29 and 30. }
\end{tabular} 
     \label{tab:tableobservation}
 \end{minipage}
\end{center}
\end{table*}

\begin{figure} 
\begin{center}
\epsfxsize=9.0cm\epsfbox{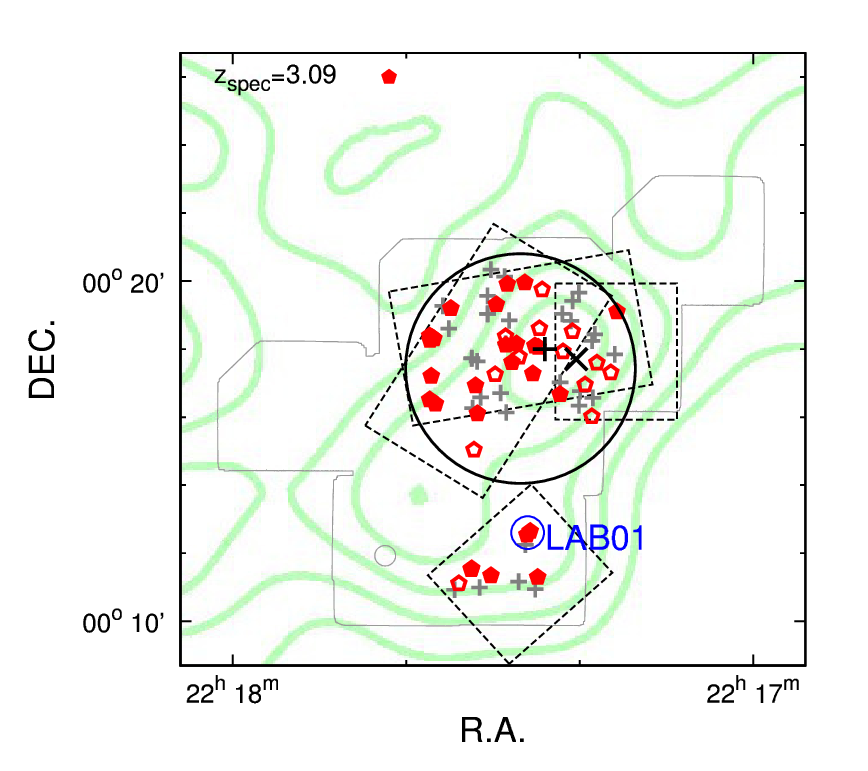}
\caption{Sky coordinates of the targets. 
The boxes with black dashed lines show the locations 
of the slit masks for our NIR spectroscopy. 
The filled red pentagons show the galaxies confirmed at $3.04\leq z_{\rm spec}\leq 3.12$ 
and the blank red pentagons show those at $z_{\rm spec}<3.04$ or $z_{\rm spec}>3.12$. 
The gray crosses show the targets observed but unidentified their redshifts.   
We use the targets within the black large circle with 1.5 Mpc radius 
in physical scale ($\approx$ 3.3 arcmin) 
to estimate the cluster mass in \S~4. 1. 
The gray solid lines enclose the FoV of the MOIRCS $K$-band images 
\citep{2012ApJ...750..116U}. 
The green contours show the surface density levels of LAEs at $z=3.09$ 
and the black large cross shows its peak obtained by \citet{2012AJ....143...79Y}.
The black large X shows the density peak of the $K$-selected galaxies 
at $2.6<z_{\rm phot}<3.6$ obtained by \citet{2013ApJ...778..170K}. 
The blue large circle indicates the location of LAB01 \citep{2000ApJ...532..170S}.  
 } \label{fig:sky-dis-targets} 
\end{center} 
\end{figure}

\begin{figure*} 
\begin{center}
\epsfxsize=18cm\epsfbox{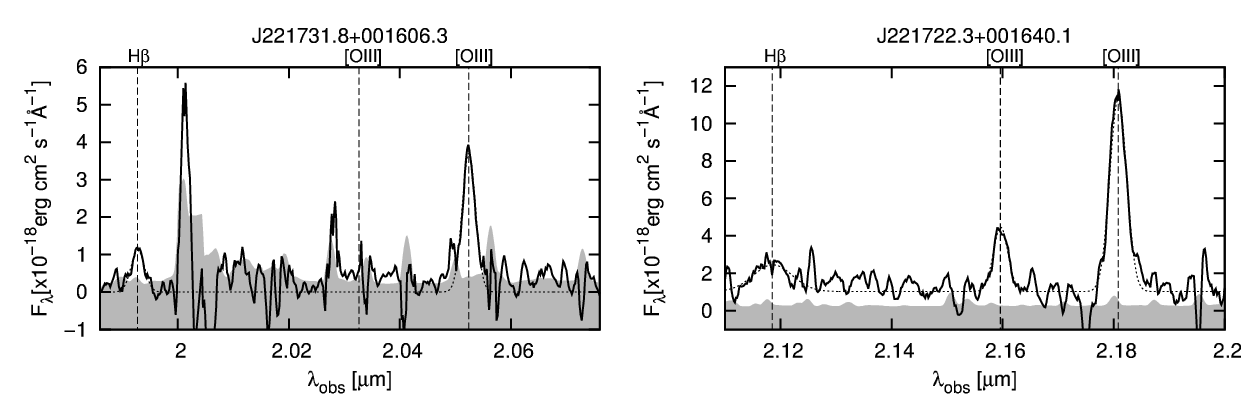}
\caption{ 
The $K$-band spectra of J221731.8+001606.3 at $z_{\rm spec}=3.0981$ ({\it left}) 
and J221722.3+001640.1 at $z_{\rm spec}=3.3544$ ({\it right}). 
The black solid lines show the obtained spectra 
and the gray shaded areas show 1$\sigma$ Poisson noise of the sky background. 
The dotted curves show the best-fit line profiles obtained by using the SPECFIT. 
The dashed vertical lines are drawn at the wavelength of [O{\scriptsize III}] $\lambda 4959,5007$ and H$\beta$ emission lines 
while the wavelength of [O{\scriptsize III}] $\lambda4959$ of J221731.8+001606.3
is shown at that expected from its [O{\scriptsize III}] $\lambda5007$ but obscured by OH airglow lines. 
} \label{fig:spec-plot} 
\end{center} 
\end{figure*}
\begin{figure} 
\begin{center}
\epsfxsize=9cm\epsfbox{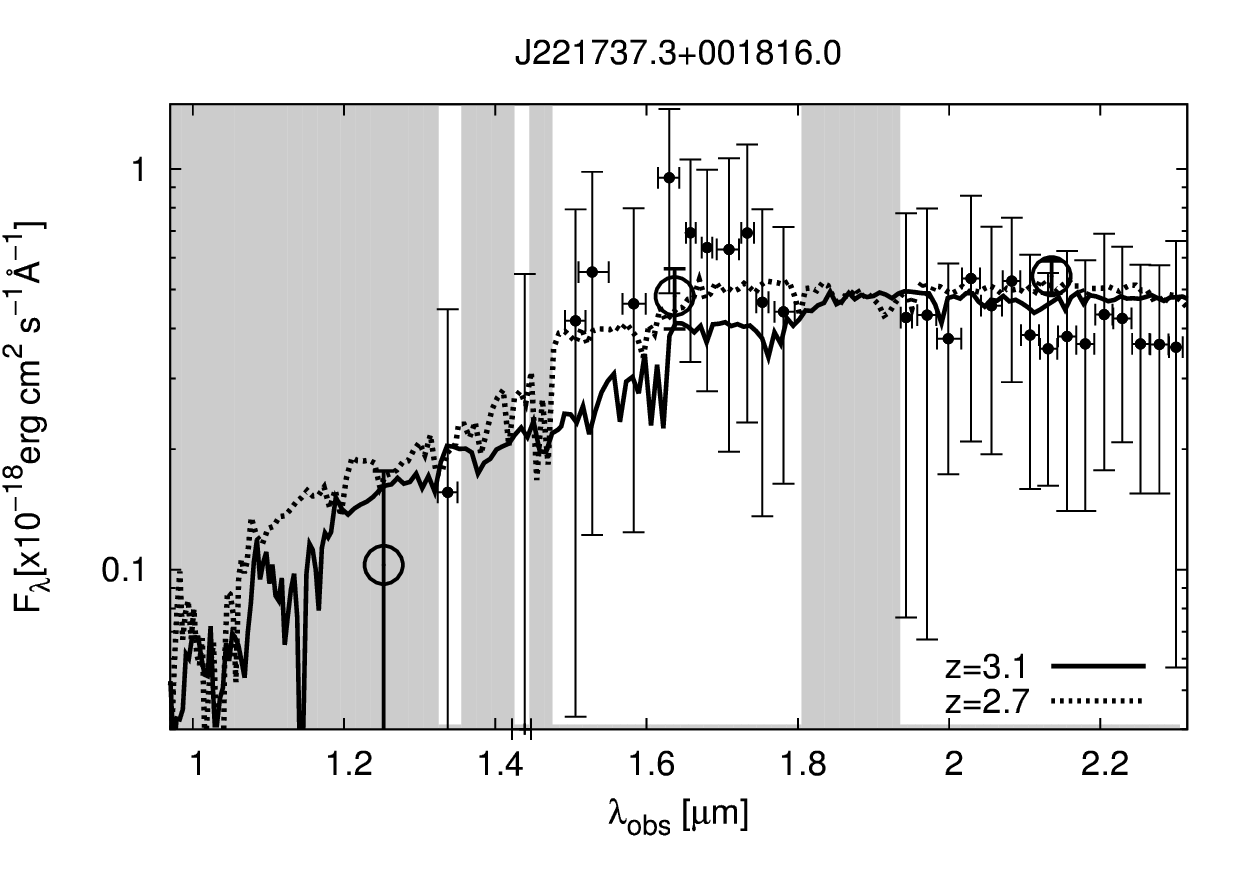}
\caption{
The stacked spectrum of J221737.3+001816.0. 
The black filled circles show the flux values stacked 
over the ranges indicated with the horizontal bars. 
The black large circles show the photometric data points 
measured on the MOIRCS $JHK$-band images. 
The black solid curve shows the best-fit SED model 
obtained with the photometric data, adopting $z=3.1$ 
($\tau=0.1$, $E(B-V)=0.1$, age $=$ 2.0 Gyr and $Z=0.2~Z_{\odot}$) 
while the full SED is shown in the {\it top} panel of Figure \ref{fig:sed-qgs}. 
The black dotted curve shows the best-fit SED model obtained adopting $z=2.7$ 
while its SED parameters are same as those obtained adopting $z=3.1$. 
The gray shaded areas indicate wavelength ranges  
obscured by atomospheric absorption and crowded OH airglow lines. 
} 
\label{fig:stackd100}
\end{center} 
\end{figure}

The observations were conducted with MOIRCS on Subaru Telescope 
during 2012 September 29-30 and October 27-28. 
Summary of the observations is given in Table \ref{tab:tableobservation}. 
We used the Multi-Object Spectroscopy (MOS) mode of MOIRCS. 
The locations of the MOS slit masks and the targets are shown in Figure \ref{fig:sky-dis-targets}.  
Our targets were the candidate members of the SSA22 protocluster
with $K<24$ and $2.6<z_{\rm phot}<3.6$, selected based on the photometric redshifts 
obtained by \citet{2012ApJ...750..116U} and \citet{2013ApJ...778..170K}. 
We used the four MOS slit masks to cover the density peaks 
of LAEs at $z\approx3.09$ \citep{2012AJ....143...79Y}
and the $K$-selected galaxies at $2.6<z_{\rm phot}<3.6$ \citep{2013ApJ...778..170K}
and also LAB01, which is known as one of the largest LABs \citep{2000ApJ...532..170S}. 
In total, we observed 67 objects while  
four of them were observed twice with the different slit masks. 

We put priority on DRGs, HEROs, {\it Spitzer} MIPS $24~\mu$m sources, 
{\it Chandra} X-ray sources, and the counterparts of LABs and the AzTEC/ASTE
1.1-mm sources (selected based on \citealt{2004AJ....128..569M, 2009ApJ...692.1561W, 2009MNRAS.400..299L, 2014MNRAS.440.3462U}). 
The numbers of the targets satisfying each classification 
are listed in Table \ref{tab:tabletargets}. 
The detection limit of the {\it Chandra} data corresponds 
to $L_{\rm X}\sim10^{43}$ erg s$^{-1}$. 
The objects with flux in MIPS $24~\mu$m-band $f_{\rm 24\mu m}\gtrsim 60~\mu$Jy 
are identified as $24~\mu$m sources. 
Such luminous mid-infrared sources at $z>2$ are not likely 
to be only originated in dusty starburst 
but also obscured AGNs (e.g., \citealt{2007ApJ...670..173D}).
Actually, of the 18 targets detected at $24~\mu$m, 
four are also detected with {\it Chandra}. 
Some of DRGs, $24~\mu$m and X-ray sources 
were selected as the targets regardless of their photometric redshifts.  

MOIRCS consists of two channels, Ch1 and Ch2, 
with each $3'.5\times4'.0$ field of view (FoV). 
We used the Volume Phase Holographic $K$-band 
grism for Ch2 (VPH-$K$; \citealt{2011PASJ...63S.605E}). 
The VPH-$K$ grism covers a wavelength range from $1.9$ to $2.3~\mu$m
with $R\sim1700$ with 0$''$.8 slit width. 
For Ch1, we had to use the $HK$500 grism due 
to the instrument trouble occurred during the run.
The $HK$500 grism covers a wavelength range from 1.3 to 2.3 $\mu$m
with $R\sim500$ with 0$''$.8 slit width. 
Ch1 could not be used for the mask alignments for the first two nights. 
The total exposure time was $3.6-4.4$ hours for each mask. 
The seeing size in the $K_{s}$-band 
during the observations was typically $0''.6$, ranging from $0''.4$ to $0''.8$. 
The telescope was dithered along the slits by $3''.0$ for sky subtractions. 
Flux standard stars for the VPH-$K$ grism were taken 
at the ends of the observations using the slits in the object masks.
The flux values were calibrated to their $K$-band total magnitudes given in  
Two Micron All Sky Survey point source catalog \citep{2006AJ....131.1163S}. 
Since acquisitions of known standard stars were not available 
for Ch1 during the run, the bright stars in the object masks 
were used for the calibration.

The data reduction was performed by using IRAF 
scripts for Subaru/MOIRCS MOS data 
(MCSMDP\footnotemark[1]) following its standard manner 
(e.g., \citealt{2010ApJ...718..112Y}). 
Note that, the order-sorting filter was not used 
during the observations on September 29 and 30  
and we corrected the contributions of the second order spectra. 
The error from this procedure is $0.4-0.7\%$ of a flux value. 

\footnotetext[1]{available at http://www.naoj.org/Observing/DataReduction/}

The $HK$500 grism provides a spectral resolution of 40\AA~with 0$''$.8 slit 
while the VPH-$K$ grism provides spectral resolutions 
of 9\AA~and 11\AA~with 0$''$.7 slit and 0$''$.8 slit, respectively.
Spectral coverages of the VPH-$K$ and $HK500$ grism 
correspond to the redshifted [O{\footnotesize III}] $\lambda$5007 emission line 
at $2.9<z<3.6$ and $1.9<z<3.6$, respectively, excluding $2.7<z<2.9$  
where spectra are buried by strong atomospheric absorption. 
Efficiency of the VPH-$K$ grism is the highest at $\sim2.05~\mu$m, 
where [O{\footnotesize III}] $\lambda$5007 at $z\approx3.09$ is shifted to. 
It basically depends on wavelength and location of a slit in the direction of dispersion  
but is almost uniform for our slit arrangements. 
Since crowded OH airglow lines are well-resolved, 
S/N ratios of the spectra obtained by using the VPH-$K$ grism 
are typically higher than those obtained 
by using the $HK500$ grism at $\sim2.05~\mu$m. 

Figure \ref{fig:spec-plot} show examples of the obtained spectra. 
The flux errors are 1$\sigma$ Poisson errors of the sky spectra before the sky subtraction at each wavelength. 
The emission lines were visually identified on the two-dimensional spectra 
and those with $\sim1-2\times10^{-17}$ erg s$^{-1}$ cm$^{-2}$ 
($\sim3\sigma$ of the background noise at $\sim2.05~\mu$m) were detectable. 
The dotted curve in each panel shows the best-fit line profile. 
The one-dimensional profiles of the spectra were inspected after combined 
the spectra within the appropriate apertures along the slit
giving the largest S/N ratios for the detected emission lines. 
We used the SPECFIT in IRAF \citep{1994adass...3..437K} 
to measure the flux values, central wavelengths 
and line width of the detected emission lines  
by fitting the multiple emission lines 
and continuum emission simultaneously. 
Emission lines were fitted with Gaussian profiles while 
continuum spectra are assumed to be linear functions of wavelength at the fitted regions. 
The obtained redshifts and flux values are listed in Table \ref{tab:tableobject}.
The errors of the best-fit profiles are shown as the errors of the redshifts and flux values. 

Spectroscopic redshifts are ideally constrained with sets of emission lines
but it often happens that only a single emission line 
is detected due to the sensitivity and obscuration by atmospheric absorption and OH airglow lines. 
We therefore also consider the redshift probability distribution 
obtained with the photometric redshift of each galaxy. 
The uncertainty of our photometric redshift is typically 
$|z_{\rm spec} - z_{\rm phot} |\sim0.5$ \citep{2013ApJ...778..170K}, 
enough accurate to distinguish galaxies at $z\sim 2$, $z\sim3$ and $z\sim4$ 
where the strongest emission lines shifted to $\sim2~\mu$m  
are H$\alpha$,  [O{\footnotesize III}] $\lambda$5007 
and [O{\footnotesize II}] $\lambda$3727, respectively. 
We also checked whether the second strongest emission line 
can be detected as a single line due to the obscuration 
of the strongest one.

One object, J221737.3+001816.0 is very likely to be at $z\approx3.09$ 
from its Balmer and 4000\AA~breaks 
but the solution of $z\sim2.7$ is also acceptable. 
Its spectrum is shown in Figure \ref{fig:stackd100}. 
The flux were stacked over the wavelength ranges indicated 
with the horizontal bar at each point avoiding the wavelength ranges 
with heavy obscuration by OH airglow lines and atomospheric absorption. 
There are significant breaks at $\sim1.4-1.5~\mu$m and $\sim1.6~\mu$m 
which are consistent with the Balmer and 4000\AA~breaks at $z=3.0-3.15$, respectively.  
On the other hand, $z\sim2.7$ is also an acceptable solution 
where the significant break at $\sim1.4-1.5~\mu$m 
corresponds to 4000\AA~break at $z\sim2.7$. 
Its whole spectrum is best-fitted with the model with 
$\tau=0.1$, $E(B-V)=0.1$, age $=$ 2.0 Gyr and $Z=0.2~Z_{\odot}$  at $z\sim3$ 
(see also \S~4.2) but an offset at $1.5$ to $1.8~\mu$m remains.  
If the offset is not simply due to the photometric errors, 
additional blue stellar components by small starbursts 
and minor mergers of bluer systems can be the plausible origins.

\begin{figure}[htbp]
\begin{center}
\epsfxsize=9cm\epsfbox{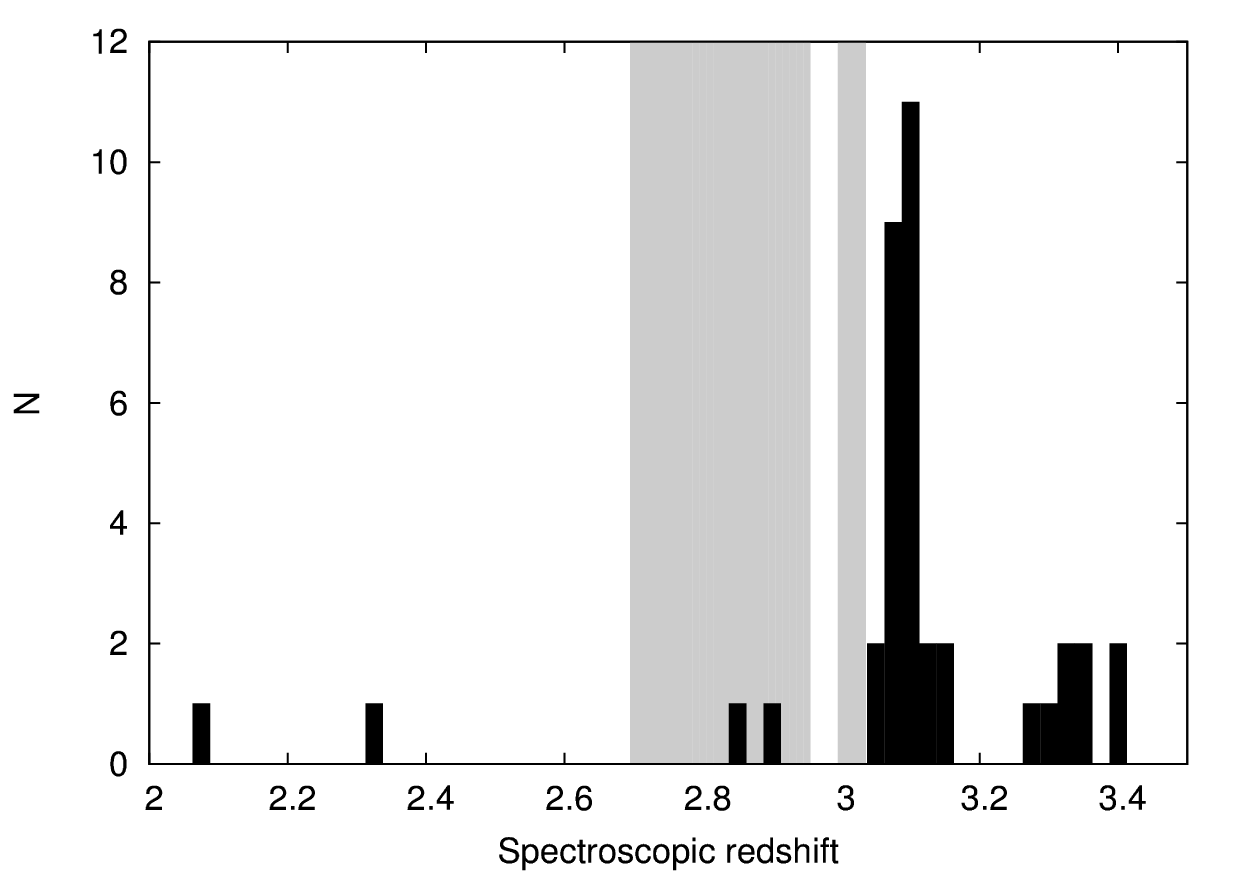}
\caption{ The redshift distribution of the galaxies in the SSA22 field 
obtained with our NIR spectroscopy. 
The black histogram shows the redshift distribution of the galaxies
confirmed with our NIR spectroscopic observations.  
The object J221737.3+001816.0 
at $z_{\rm spec}\sim2.7, 3.0-3.15$  is not included here. 
At $z > 2.7$, the gray shaded areas indicate redshift ranges 
where it is hard to confirm redshifts 
from the redshifted [O{\scriptsize III}] $\lambda5007$ due 
to atomospheric absorption and crowded OH airglow lines. 
 } 
\label{fig:dis-targets} 
\end{center} 
\end{figure}

\section{Results}
Table \ref{tab:tableobject} is the list of the 39 galaxies whose redshifts were obtained.  
Figure \ref{fig:dis-targets} shows the redshift distribution of them. 
Note that it is hard to observe [O{\footnotesize III}] $\lambda$5007 
at $z=3.00-3.03$ and $z=2.7-2.95$ (gray shaded region) 
due to atomospheric absorption and crowded OH airglow lines. 
There is a very clear redshift spike at $z\approx3.09$; 
Of the 39 galaxies, 24 are at $3.04\leq z_{\rm spec}\leq 3.12$, 
which corresponds to a relative radial velocity of $\pm$ 3000 km s$^{-1}$, 
and considered to be the protocluster members. 
As we describe in the next section, the velocity dispersion of the SSA22 protocluster 
is $\approx 1000$ km s$^{-1}$ and 
the $\pm~3\sigma$ range is regarded as the cluster members. 

Of the 39 galaxies, 11 were known to be at $2.8<z_{\rm spec}<3.4$ 
from previous optical and NIR spectroscopic observations 
\citep{2001ApJ...554..981P, 2003ApJ...592..728S, 2006ApJ...651..688S,2009MNRAS.400..299L, 2012AJ....143...79Y}. 
Our spectroscopic redshifts agree well with those in literatures 
while the largest redshift offset is $\Delta z=0.07$ for J221725.4+001716.9 
at $z_{\rm H\beta}=3.0482$ ($z_{\rm lit}=3.120$).  

Stellar masses of the galaxies were estimated based on the SED fitting 
of the fluxes in the $u^{\star}BVRi'z'JHK$, 3.6, 4.5, 5.8 and $8.0~\mu$m -bands 
\citep{2004AJ....128.2073H, 2004AJ....128..569M, 2009ApJ...692.1561W, 2012ApJ...750..116U}
through a standard $\chi^2$ minimization procedure. 
The detail of the procedure was described in \citet{2013ApJ...778..170K}.  
Briefly, we fitted the observed flux values with the \citet{2003MNRAS.344.1000B} models 
with exponentially decay star formation histories 
where SFR $\propto \exp (-t/\tau)$ $(\tau=0.1- 30$ Gyr), 
metallicity $Z=0.005, 0.02, 0.2,  0.4, 1 ~Z_{\odot}$ 
and the \citet{2000ApJ...533..682C} extinction law ranging $E(B-V)=0.0-2.0$.  
We here adopt the Chabrier Initial Mass Function (IMF) \citep{2003PASP..115..763C}
while we used the Salpeter IMF \citep{1955ApJ...121..161S} in \citet{2013ApJ...778..170K}. 
Before we perform the SED fittings, 
we applied slight corrections for the [O{\footnotesize III}] $\lambda$5007 line fluxes 
to the $K$-band fluxes by using the observed values. 
The errors of the stellar masses show the 68\% confidence level of the probability distribution 
for the stellar mass calculated from a minimum $\chi^2$ value for each object. 

Many LBGs and LAEs have already been 
identified as the members of the SSA22 protocluster 
with the observations at optical wavelength 
\citep{1998ApJ...492..428S, 2000ApJ...532..170S, 2004AJ....128.2073H, 2012AJ....143...79Y}. 
Our $K$-band selected and confirmed protocluster galaxies are typically 
faint in rest-frame UV; only five are detected in the existing 
Subaru $V$-band image \citep{2004AJ....128.2073H};  
Ly$\alpha$ are detected in the narrow-band image 
($NB497$; \citealt{2004AJ....128.2073H}) 
above 4$\sigma$ significance for eleven. 
The median stellar mass of the protocluster galaxies in our sample 
is $\simeq5\times10^{10}~M_{\odot}$ while that 
of the protocluster DRGs is  $\simeq8\times10^{10}~M_{\odot}$. 
The typical stellar mass of LBGs at $z=2-3$ is 
$\simeq6\times10^9 ~M_{\odot}$ (mean value; \citealt{2001ApJ...559..620P})
and those with $K_s <22.5$ in Vega ($\lesssim$ 24.3 in AB)
is $2.5\times10^{10}~M_{\odot}$ (median value; \citealt{2001ApJ...562...95S}), 
adopting solar metallicity and the Salpeter IMF  
while stellar mass estimated with the Salpeter IMF is 
about twice of that estimated with the Chabrier IMF on average 
(e.g., \citealt{2006ApJ...646..107E, 2010ApJ...718..112Y}). 
From these above, we emphasize that we identified 
the galaxy population different from those previously 
found in the SSA22 protocluster. 

We summarize the results of the spectroscopy of DRGs, HEROs, 
$24~\mu$m sources, X-ray sources  and 
LAEs \citep{2004AJ....128.2073H} in Table \ref{tab:tabletargets}. 
Many of them are certainly the members of the SSA22 protocluster. 
We also observed the candidate counterparts of the 
four LABs \citep{2004AJ....128..569M} and the three AzTEC/ASTE 1.1-mm sources 
\citep{2014MNRAS.440.3462U}. 
More than one counterparts at $z_{\rm spec}\approx3.09$ 
were identified for each object excluding one 1.1-mm source 
(AzTEC01 in \citealt{2014MNRAS.440.3462U}). 
The presence of such multiple counterparts for many LABs and SMGs 
in the SSA22 field were reported in \citet{2012ApJ...750..116U}.
They are likely to be the hierarchical multiple mergers 
at the early-phase of the formation history of massive early-type galaxies,  
predicted in the cosmological numerical simulations
(e.g., \citealt{2003ApJ...590..619M, 2007ApJ...658..710N, 2010ApJ...725.2312O}). 

Spatial extents of [O{\footnotesize III}] $\lambda$5007 emission lines $r_{[{\rm O}_{\rm III}]}$ 
of the galaxies were measured with FWHM of the emission lines along 
the slits after deconvolved with the PSF profiles.
The average values and standard deviations of $r_{[{\rm O}_{\rm III}]}$ 
of the protocluster galaxies and the field galaxies 
are $5.0\pm1.1$ kpc and $6.0\pm1.4$ kpc, respectively. 
Velocity dispersions $\sigma_v$ of the galaxies were measured for those observed 
by using the VPH-$K$ grism since its instrumental resolution
is small enough to resolve kinematics 
of galaxies with $\sigma_v\gtrsim70$ km s$^{-1}$. 
The average and standard deviation of the velocity dispersion 
of our sample are $137\pm50$ km s$^{-1}$ 
and $112\pm51$ km s$^{-1}$ for the protocluster galaxies 
and the field galaxies, respectively, 
excluding the outliers with $\sigma_v>300$ km s$^{-1}$. 
The velocity dispersions of the protocluster galaxies are 
slightly larger on average but do not differ 
significantly from those in the field in our sample.
The spatial extents and velocity dispersions of our sample 
are similar to those of the field galaxies at $z\sim2$ 
with similar stellar mass \citep{2006ApJ...646..107E} 
but larger than those of LBGs at $z=2-3$,  
which have half light radii $0''.1-0''.4$, corresponding to $0.8\sim3$ kpc, 
at rest-frame optical wavelength
(e.g., \citealt{1996ApJ...470..189G, 2008ApJS..175....1A})
and $\sigma_v \lesssim120$ km s$^{-1}$ measured with [O{\footnotesize III}] $\lambda$5007 
(\citealt{2001ApJ...554..981P, 2011A&A...528A..88G}, 
maybe biased to bright LBGs). 
This also supports that the $K$-selected members are 
the relatively massive galaxy population in the SSA22 protocluster, 
since [O{\footnotesize III}] line widths and spatial extents 
depend on dynamical properties of gas and stellar components of galaxies. 

\begin{table}
 \caption{Summary of the targets}
 \begin{center}
\begin{tabular}{lccc}
\hline \hline
Classification & N$_{\rm targets}$ & N$_{\rm cluster}$\footnotemark[a]  & N$_{\rm field}$\footnotemark[b] 
\\ \hline
All & 67 & 24 & 15 \\
$K<24$ \& $2.6<z_{\rm phot}<3.6$ & 56 & 21 & 12 \\
DRGs ($J-K>1.4$) & 21 & 11 & 4 \\
HEROs ($J-K>2.1$) & 9 & 6 & 1 \\
$24~\mu$m sources & 18 & 7 & 5 \\
X-ray sources & 14 & 4 & 4 \\
LAEs & 5(12)\footnotemark[c] & 5(11)\footnotemark[c] & 0 \\
\hline
\footnotetext[a]{Number of the galaxies confirmed the redshifts at $3.04\leq z_{\rm spec} \leq 3.12$. }
\footnotetext[b]{Number of the galaxies confirmed the redshifts at $z<3.04$ and $z>3.12$.}
\footnotetext[c]{Number of the LAEs selected with the narrow-band filter $NB497$ 
which is sensitive to Ly$\alpha$ emission line at $z=3.062-3.125$ 
\citep{2004AJ....128.2073H, 2012AJ....143...79Y}. 
The numbers in the brackets show the number of the galaxies with 
the $BV-NB497$ color excess above 4$\sigma$ significance.  }
\end{tabular} 
     \label{tab:tabletargets}
\end{center}
\end{table}

\section{Discussion}
\subsection{The mass of the SSA22 protocluster}
First, we discuss the mass of the SSA22 protocluster itself. 
Again, our NIR spectroscopic observations were conducted 
to cover the central highest density region of LAEs at $z=3.09$ 
extended to $\sim 3\times 6$ Mpc in physical scale 
\citep{2012AJ....143...79Y}, which may evolve into a cluster core. 
Consequently, the region within $1.5$ Mpc in physical 
($\approx3.3$ arcmin) from the density peak,  
indicated with the black large circle in Figure \ref{fig:sky-dis-targets}, 
was observed almost uniformly. 
This region is also uniformly covered with optical spectroscopic observations 
\citep{2003ApJ...592..728S, 2009MNRAS.400..299L, 2005ApJ...634L.125M, 2012AJ....143...79Y}. 
We here estimate the mass enclosed within this region. 
Note that the entire high density region of the SSA22 superstructure 
may extend to $\sim100$ Mpc in the comoving scale 
\citep{2012AJ....143...79Y}. 

\begin{figure}[h]
\begin{center}
\epsfxsize=9.0cm\epsfbox{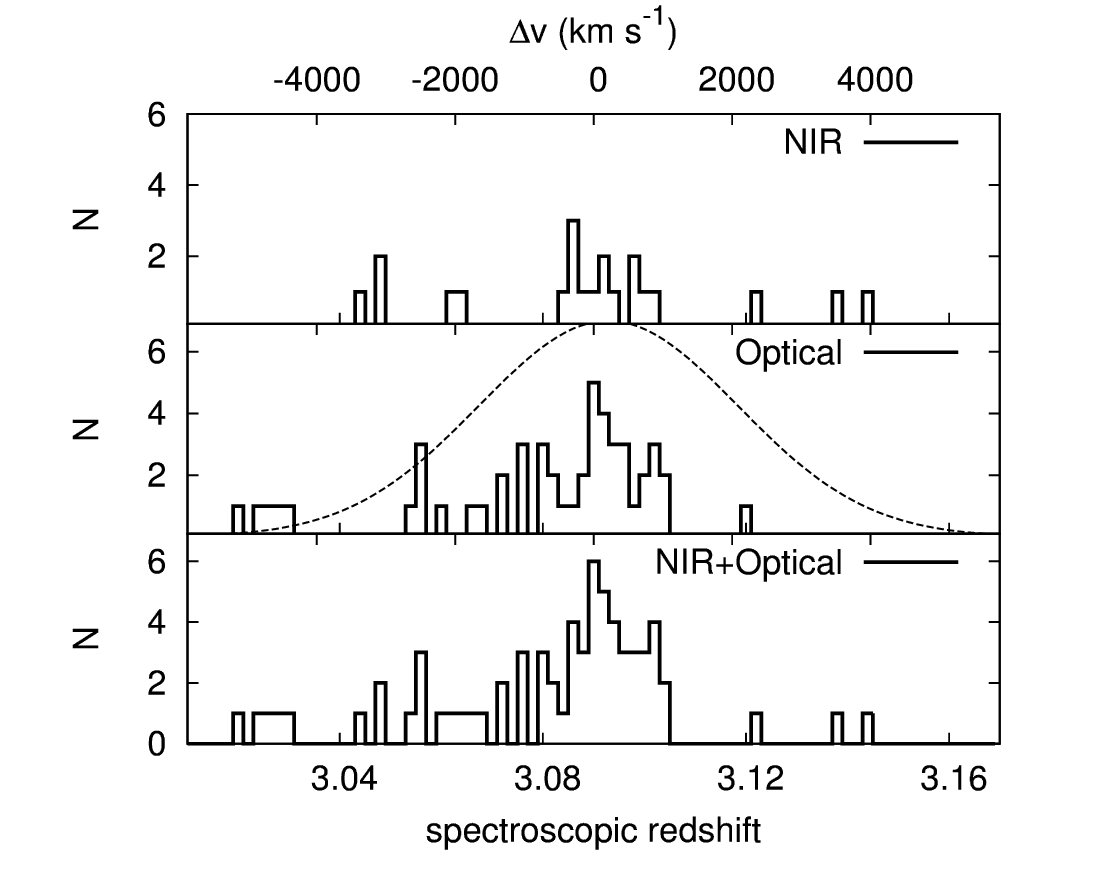}
\caption{The redshift distributions of the galaxies 
within the central 1.5 Mpc of the SSA22 protocluster 
shown in Figure \ref{fig:sky-dis-targets}. 
The histograms show the redshift distributions of our sample ({\it top}), 
the galaxies confirmed the redshifts with optical spectroscopic observations in literatures
({\it middle}; based on \citealt{2003ApJ...592..728S, 2009MNRAS.400..299L}; 
\citealt{2005ApJ...634L.125M, 2012AJ....143...79Y})
and the combined sample  ({\it bottom}). 
The {\it top} axis shows the velocity scaled taking $z=3.09$ as a center. 
The dashed curve in the {\it middle} panel shows 
the transmission curve of the $NB497$ filter, 
namely the selection function for LAEs at $z=3.09$.
 } \label{fig:dis-targets-cent} 
\end{center} 
\end{figure}

Figure \ref{fig:dis-targets-cent} shows the redshift distributions 
of the protocluster galaxies within this region. 
The numbers of the galaxies at $3.04\leq z_{\rm spec}\leq 3.12$ within this radius 
are 18 for our sample, 45 for the sample confirmed with optical spectroscopic observations 
\citep{2003ApJ...592..728S, 2009MNRAS.400..299L, 2005ApJ...634L.125M, 2012AJ....143...79Y}
and 60 for the combined sample where we use our own redshifts for the overlapping objects. 
The redshift selection functions for LBGs and the photo-z selected galaxies 
are uniform at a redshift range from $z=3.04$ to 3.12 
while LAEs were nonuniformly selected in $z=3.062-3.125$ 
and preferentially picked up around $z\approx3.09$ due to the narrow-band selection  
shown in the {\it middle} panel of Figure \ref{fig:dis-targets-cent}. 

It was reported that the redshift obtained with 
Ly$\alpha$ offsets for several 100 km s$^{-1}$ from the systemic one  
obtained with nebular emission lines 
\citep{2010ApJ...717..289S, 2011ApJ...729..140F, 2011ApJ...730..136M, 2014ApJ...795...33E}. 
It is thought that this offset of Ly$\alpha$ is caused 
by resonant scatterings in H{\footnotesize I} outflowing gas.  
The Ly$\alpha$ redshifts of our sample
redshift by $40\sim640$ km s$^{-1}$ from our own [O{\footnotesize III}] redshifts. 
We here ignore the offsets in the redshifts obtained with Ly$\alpha$ 
in calculating the velocity dispersion of the protocluster.

As seen in Figure \ref{fig:dis-targets-cent}, 
the galaxies in each sample are clearly clustered 
at the redshift ranges narrower than those expected 
from the redshift selection function of each sample. 
By fitting each of the redshift distributions at $3.04\leq z \leq 3.12$ with a Gaussian profile, 
we obtain the velocity dispersions $\sigma_v = 521\pm122$ km s$^{-1}$, 
$790\pm127$ km s$^{-1}$ and $658\pm84$ km s$^{-1}$ 
for the NIR, optical and combined samples, respectively. 
The centers of the redshift distributions are at $z=3.091\sim3.092$
with the errors $\Delta z=0.001\sim0.002$, agree well each other. 
If we simply evaluate the $\sigma_v$ as the standard deviation 
of each redshift distribution at $3.04\leq z \leq 3.12$, 
we obtain $\sigma_v=1394$, 1100 and 1174 km s$^{-1}$ 
for the NIR, optical and combined samples, respectively. 

The velocity dispersions of our samples are similar to those 
of the protoclusters at $2<z<4$ 
obtained from LAEs by \citet{2007A&A...461..823V} but 
higher than those of $z=3$ progenitors of massive clusters 
with $\sigma_v\sim1000$ km s$^{-1}$ in the local Universe, 
predicted from the numerical simulations (e.g., \citealt{1998ApJ...503..569E}).
If we assume the virial equilibrium, the virial mass of this protocluster is 
$M_{\rm vir}\sim3{\sigma_v}^2 r/G =4.5\pm1.1\times10^{14}$ 
$M_{\odot}$ $(\sigma_{v}/658$ km s$^{-1})^2(r/1.5$ Mpc$)$, 
where $G$ is the gravitational constant. 

It is, however, not likely to happen that the SSA22 protocluster 
has already been fully virialized. 
We then also evaluate the mass of the protocluster 
from its overdensity following \citet{1998ApJ...492..428S} 
and \citet{2005A&A...431..793V}. 
The mass of a protocluster with comoving volume $V$ 
and mass overdensity $\delta_{\rm m}$, 
which is estimated from that of galaxies, 
is given as $M=\bar{\rho}V (1+\delta_{\rm m})$ 
where $\bar{\rho}$ is the mean density of the Universe. 
The mean density $\bar{\rho}$ is $\approx4.2\times10^{10}~M_{\odot}$ Mpc$^{-3}$ at $z=3.09$ 
for the cosmological parameters adopted in this paper. 
We here take the volume within the area same as 
that used to evaluate the velocity dispersion 
and the redshift range $3.08<z<3.10$ ($\approx1500$ km s$^{-1}$),  
which corresponds to $V\approx2.2\times10^3$ Mpc$^3$, 
as the comoving volume of the protocluster.

Overdensity of the galaxies in the protocluster is defined as the ratio 
of the number density of the galaxies in the protocluster ($n_{\rm cluster}$), 
to that in the field ($n_{\rm field}$) as $\delta_{\rm gal}=(n_{\rm cluster}/n_{\rm field}-1)$. 
We here estimate the overdensity from the fraction of the galaxies in the SSA22 protocluster 
among the targets with $K<24$ and $2.6<z_{\rm phot}<3.6$. 
Of the targets with $K<24$ and $2.6<z_{\rm phot}<3.6$ 
within the central 3.3 arcmin radius, 12 galaxies were confirmed at $3.08 < z < 3.10$ 
while the rest 32 galaxies were not confirmed at $3.04 \leq z \leq 3.12$. 
Assuming that the latter galaxies are the field galaxies 
within the redshift range of $2.6<z<3.6$, 
the overdensity of the galaxies in the SSA22 protocluster is estimated 
to be $\delta_{\rm gal}=17.8\pm5.4$. 
This is a lower limit value since not all the protocluster galaxies may be 
enough bright to be identified the redshifts by the detections of [O{\footnotesize III}]. 

Following \citet{1998ApJ...492..428S}, 
the mass overdensity $\delta_{\rm m}$ is related to the $\delta_{\rm gal}$ 
as $1+b\delta_{\rm m}=C(1+\delta_{\rm gal})$, taking into account 
for the redshift space distortion caused by peculiar velocities. 
$C$ can be approximated by $C=1+f-f(1+\delta_{\rm m})^{1/3}$.  
$f$ is the growth rate of the perturbations at the given redshift 
\citep{1991MNRAS.251..128L}, 
$\sim1$ at $z>2$ for the cosmology adopted here. 
Adopting the bias parameter $b=3\sim5$  
(e.g., \citealt{2007ApJ...654..138Q, 2007PASJ...59.1081I}),   
we obtain the $\delta_{\rm m}=1.9\sim2.6$,  
which corresponds to the cluster mass of $1.8\sim2.4\times10^{14} ~M_{\odot}$. 

\begin{figure}[!t]
\begin{center}
\epsfxsize=9cm\epsfbox{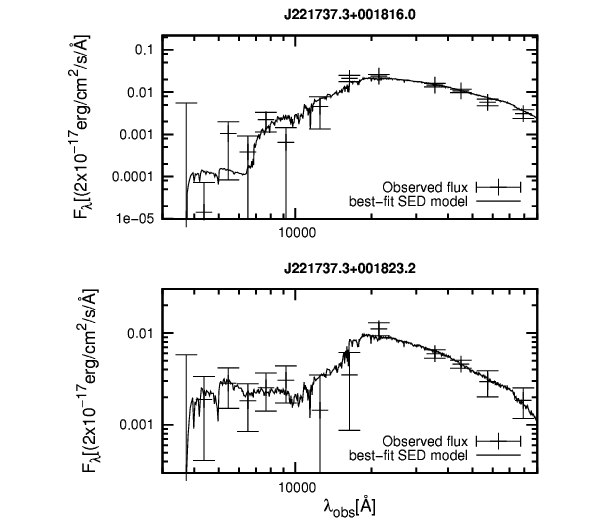}
\caption{ 
Examples of the SEDs of the massive quiescent 
galaxies confirmed in the SSA22 protocluster. 
The cross points are the observed flux values in the $u^{\star}BVRi'z'JHK$, 
3.6, 4.5, 5.8 and 8.0 $\mu$m -bands. 
The solid curves indicate their best-fit SED models.
The best-fit SED models of J221737.3+001816.0 ({\it Top}) 
and J221737.3+001823.2 ({\it Bottom}) 
are those with SFR $\propto \exp (-t/\tau)$ where $\tau=0.1$, 
$E(B-V)$=0.1, age = 2.0 Gyr and $Z=0.2~Z_{\odot}$, 
and $\tau=0.3$, $E(B-V)$=0.1, age = 0.7 Gyr and $Z=0.4~Z_{\odot}$, respectively. 
 } \label{fig:sed-qgs} 
\epsfxsize=9cm\epsfbox{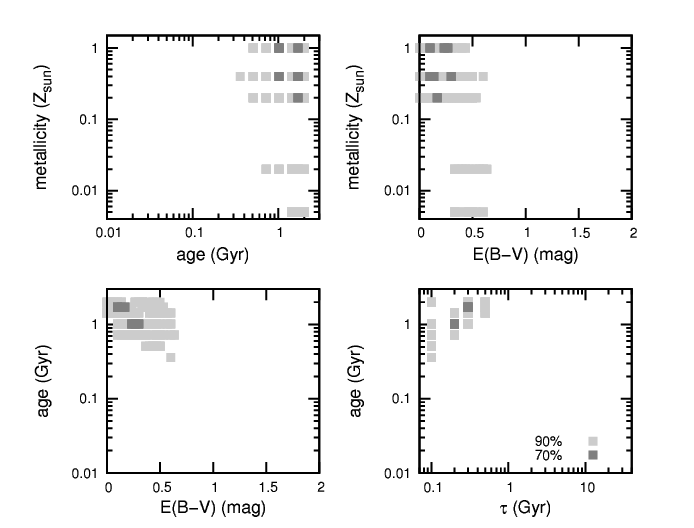}
\caption{ 
The stacked probability distributions of age, metallicity, $\tau$
and $E(B-V)$ parameters of all the protocluster galaxies with $J-K\geq 2.4$. 
The dark gray and gray shaded areas show the 70\% 
and 90\% confidence regions, respectively. 
} \label{fig:sed-prob} 
\end{center} 
\end{figure}

Both methods found a cluster mass of several $10^{14} ~M_{\odot}$, 
suggesting that the mass of the SSA22 protocluster is already comparable to 
those of massive clusters in the local Universe. 
Such a massive halo at $z=3$ is very rare, may be unique  
in the horizon, as predicted from the cosmological 
numerical simulations in the $\Lambda$CDM universe 
(e.g., \citealt{2002MNRAS.336..112M}). 
Since the SSA22 protocluster is an extremely 
outstanding structure at $z=3$
as ever, it can happen that this protocluster 
is actually the one very rare cluster,  
although we should take carefully this result.   
We should note that our mass estimate is a conservative value 
since we adopt the volume including only the central part 
of the protocluster and its overdensity can be underestimated. 

\subsection{The red galaxies in the protocluster} 

Significant fractions of DRGs and HEROs in the SSA22 field 
were found to be the members of the protocluster.   
This spectroscopically supports our previous studies 
that reported the density excess of such red galaxies in the SSA22 protocluster
\citep{2012ApJ...750..116U, 2013ApJ...778..170K}. 

We found that some of these red galaxies are quiescent galaxies 
which dominate the reddest color range. 
Of the six galaxies with $J-K\geq 2.4$ observed in this study, 
five were confirmed the redshifts at $z\approx3.09$ while there are as a whole 13 
such objects found in our photometric sample in the SSA22 field. 
Figure \ref{fig:sed-qgs} shows examples of the SEDs 
of the protocluster galaxies with $J-K \geq 2.4$ 
and Figure \ref{fig:sed-prob} shows the stacked corresponding 
probability distributions of age, $\tau$, dust extinction 
and metallicity parameters of all of them. 
Four galaxies are fitted with the $\chi^2/\nu<0.84$  
while the $\chi^2/\nu=1.74$ for the rest one (J221737.3+001816.0, described in \S~2) 
due to the large error of its rest-frame UV flux. 
They show quiescent star formation activities 
and are likely to be dominated by old stellar populations 
while their best-fit SED parameters are
$\tau=0.1-0.3$ Gyr, age $=0.7-2.0$ Gyr, 
$Z =0.2-1~Z_{\odot} $, $M_{\rm star}=10^{10.9}-10^{11.5}$ $M_{\odot}$ 
and specific SFR (sSFR) $<0.1$ Gyr$^{-1}$. 
As for the SED models within the 70\% confidence range of the stacked probability distribution,  
more than $90\%$ of stars in an object are older than 0.7 Gyr. 
Although the SED models within the 90\% confidence range 
have the ages ranging from 0.3 to 2.0 Gyr, their sSFR are smaller than $0.4$ Gyr$^{-1}$. 
As we describe below, four of them are AGNs detected with {\it Chandra}. 
The origin of their [O{\footnotesize III}] $\lambda5007$ emission lines 
should be AGNs but there is no AGN feature in their rest-frame UV to NIR SEDs.  
They may be buried in the stellar light as is often the case 
for AGNs at high redshift (e.g., \citealt{2009ApJ...699.1354Y}). 
The other one without {\it Chandra} detection is J221737.3+001816.0, 
confirmed from the Balmer and 4000\AA~breaks. 
High AGN fractions in massive quiescent galaxies at $z=2-3$
were also reported in other studies 
(e.g., \citealt{2013PASJ...65...17T, 2014ApJ...783L..14S}). 

From these above, they are well-characterized 
as massive galaxies with quenched star formation. 
This may be the first time to confirm the association of a bunch of massive quiescent 
galaxies in a protocluster at $z>3$, spectroscopically. 
This also confirms our previous results that reported the overdensity 
of massive quiescent galaxies in the SSA22 protocluster 
selected based on the photometric redshifts and the color selections \citep{2013ApJ...778..170K}. 
They are very likely to take a part of progenitors of massive early-type galaxies 
at cores of massive clusters in the current Universe. 

Of the five quiescent galaxies, four are AGNs detected with {\it Chandra}. 
Their X-ray luminosities at $2-10$ keV 
evaluated from their hard band ($2-8$ keV) flux values  
assuming the spectra in form $f_{\nu}\sim\nu^{-0.7}$ 
are $L_{\rm X}\sim3-7\times10^{43}$ ergs s$^{-1}$. 
They have the X-ray luminosities lower than those of the other X-ray sources associated 
with LAEs and LBGs in the SSA22 protocluster \citep{2009ApJ...691..687L}. 
Specific AGN activities, X-ray luminosities divided 
with stellar masses of host galaxies, of them are   
$L_{\rm X}/M_{\rm star}=3-8\times10^{32}$ (erg s$^{-1}$ ${M_{\odot}}^{-1}$), 
calculated following \citet{2009ApJ...699.1354Y}.  
These values are similar to or slightly higher than those of local massive galaxies, 
if we assume the similar black hole to stellar mass ratio. 
It suggests that the massive quiescent galaxies in the SSA22 protocluster 
may be at the phase after the peaks of not only the star formation activities 
but also the mass accretion into their central black holes.
We can consider two scenarios for the declines of both the activities; 
quench of mass accretion to the central black holes due to exhaustion of the gas 
by star formation and/or quench of star formation 
by feedback from the AGNs. 
The latter is an important scenario in the recent cosmological numerical simulations 
since this may solve the over cooling problem at the high-mass end 
(e.g., \citealt{2005MNRAS.356.1155C, 2006MNRAS.370..645B}). 
 
We also identified the six DRGs with $1.4<J-K<2.4$ as the protocluster members. 
They are too faint to constrain the robust SED properties but 
their [O{\footnotesize III}] $\lambda5007$ may be originated in the 
star formation activities since there is no significant AGN feature. 
At least, one of them is likely to be a heavily dust obscured galaxy 
since it is detected at $24~\mu$m 
and also located within the beam size of an 1.1-mm source, AzTEC 99 
\citep{2014MNRAS.440.3462U}. 

\subsection{Emission line properties}

As we reported in \S~3, there is no significant difference 
in widths and spatial extents of emission lines 
between the $K$-selected galaxies in the protocluster and field whereas 
their other properties distinctly differ from 
those of rest-frame UV selected galaxies such as LAEs and LBGs. 

\begin{figure} 
\begin{center}
\epsfxsize=8.5cm\epsfbox{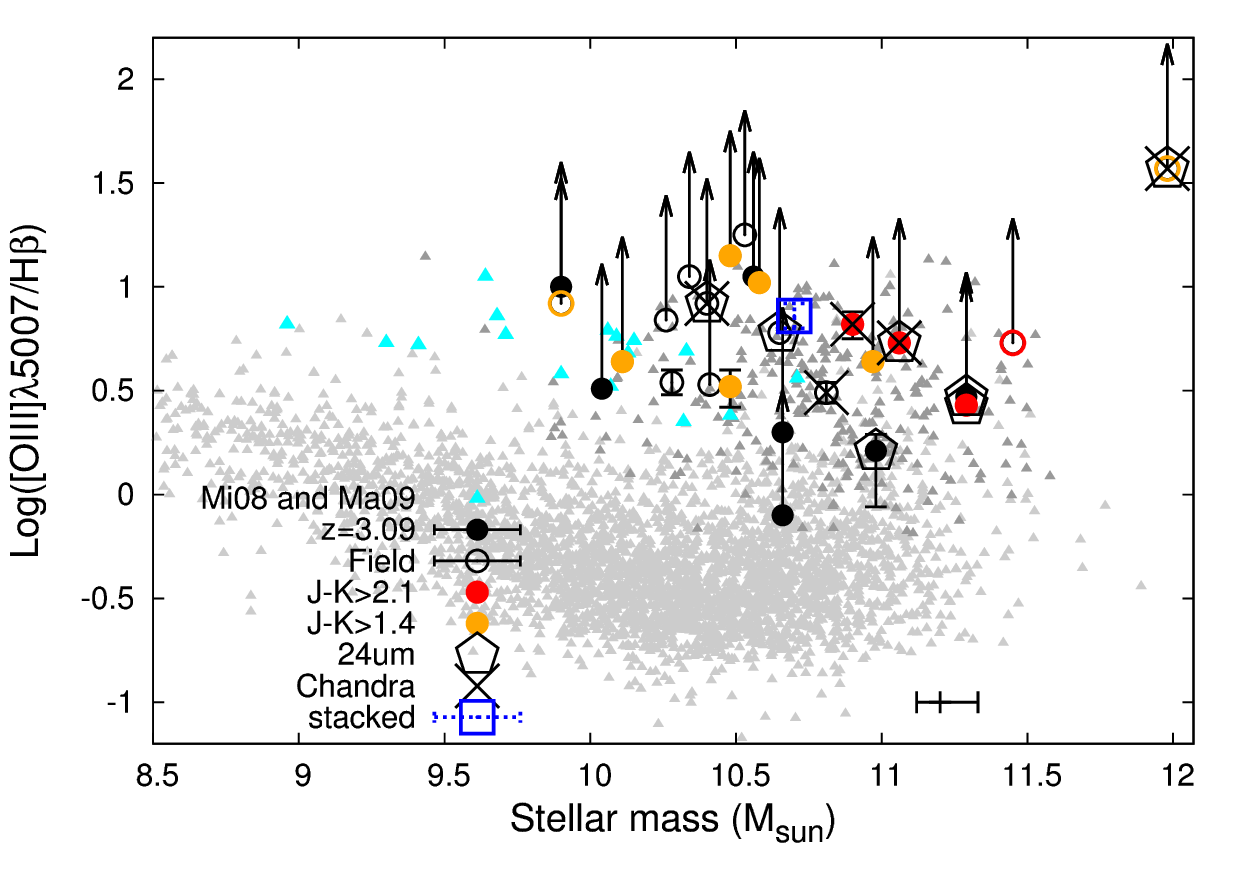}
\caption{[O{\scriptsize III}] $\lambda$5007/H$\beta$ ratio 
versus stellar mass diagram of our sample. 
The filled circles show the galaxies at $3.04\leq z_{\rm spec} \leq 3.12$ and 
the blank circles show the field galaxies at $2.9<z_{\rm spec}<3.4$ in our sample. 
The galaxies classified as $J-K\leq 1.4$, $1.4<J-K\leq 2.1$ and $J-K < 2.1$, 
are shown with the black, orange and red points, respectively. 
The blue square shows the stacked value of the galaxies confirmed 
by the detections of [O{\scriptsize III}] $\lambda 5007$ but without detections of H$\beta$ in our sample. 
The large pentagons and crosses mark the $24~\mu$m and X-ray sources, respectively.
The light and dark gray points show the galaxies 
and AGNs in the SDSS catalog \citep{2004ApJ...613..898T}. 
The cyan triangles show the LBGs at $z\sim3$ 
\citep{2008A&A...488..463M, 2009MNRAS.398.1915M}.
 } \label{fig:sm-oiii} 
\end{center} 
\end{figure}

Figure \ref{fig:sm-oiii} shows the [O{\footnotesize III}] $\lambda$5007/H$\beta$ 
line flux ratio versus stellar mass distribution, 
called Mass Excitation (MEx) diagram 
\citep{2011ApJ...736..104J}, of our sample. 
Lower limit values of the [O{\footnotesize III}] $\lambda$5007/H$\beta$ ratios 
are shown for the galaxies without detection of H$\beta$. 
The blue square shows the stacked value of all the galaxies confirmed 
by the detections of [O{\footnotesize III}] $\lambda5007$ but without detections of H$\beta$. 
[O{\footnotesize III}] $\lambda$5007/H$\beta$ ratios 
obtained with the $HK500$ grism are not shown  
since they are contaminated by OH airglow lines. 
We also plot the star-forming galaxies and AGNs in the SDSS catalog 
(using the catalog provided by \citealt{2004ApJ...613..898T}) 
and the LBGs at $z\sim3$ studied by 
\citet{2008A&A...488..463M} and \citet{2009MNRAS.398.1915M} for comparison. 

[O{\footnotesize III}] $\lambda$5007/H$\beta$ ratios 
of galaxies at $z=2-3$ are typically higher 
than those of local star-forming galaxies  
(e.g., \citealt{2006ApJ...644..813E, 2008A&A...488..463M, 2009MNRAS.398.1915M, 2011ApJ...743..144T}).  
It is thought to originate in high SFRs and low metallicities of high redshift galaxies.  
An AGN can also enhance [O{\footnotesize III}] $\lambda$5007/H$\beta$ ratio 
but it may not be so common as found from the emission line diagnostics of 
$z\sim2$ galaxies (e.g., \citealt{2006ApJ...644..813E}). 
The galaxies in our sample show large [O{\footnotesize III}] 
$\lambda$5007/H$\beta$ ratios on average, similar to LBGs at $z\sim3$. 
The AGNs in the protocluster show large [O{\footnotesize III}] $\lambda$5007/H$\beta$ ratios 
and stellar masses similar to those of local AGNs 
and also AGNs at $z\sim1$ \citep{2011ApJ...736..104J}. 

One counterpart of LAB01 has 
Log ([O{\footnotesize III}] $\lambda$5007/H$\beta)=0.21^{+0.29}_{-0.06}$, 
which is lower than those of other star-forming galaxies at $z\sim3$. 
This galaxy is an active starburst galaxy 
detected at $24~\mu$m and its SED is well-fitted 
with a dusty starburst galaxy model. 
Then its ionizing radiation field may be similar to 
or harder than those of other star-forming galaxies. 
One possible interpretation is that this galaxy has reached the level 
of chemical enrichment of galaxies in the local Universe, 
as [O{\footnotesize III}]/H$\beta$ ratio depends 
on metallicity at the given ionization parameter. 
The difference in the chemical enrichment in the protocluster 
was reported in \citet{2013ApJ...774..130K} that 
the galaxies in the protocluster at $z\sim2$ have various metallicities 
regardless of the stellar masses whereas field galaxies follow the well-established trend 
in the mass-metallicity relation at that epoch. 
At this point, it is hard to constrain the metallicities of the protocluster galaxies 
based on our spectroscopic data and also SED fittings of the photometric data. 
Further spectroscopic observations for [O{\footnotesize II}] $\lambda 3727$ emission lines  
are required to constrain the metallicities of the galaxies in the SSA22 protocluster. 

\section{Conclusion}
We conducted NIR spectroscopy of the rest-frame optically-selected candidate 
members of the protocluster at $z=3.09$ in the SSA22 field. 
We observed the 67 candidates and confirmed the redshifts of the 39 galaxies. 
We identified the 24 protocluster galaxies at $3.04\leq z_{\rm spec} \leq 3.12$ 
which are on average more massive than those identified 
by the previous observations in rest-frame UV. 
This may be one of the first observations to confirm a large number 
of galaxies at $z\sim3$ from the redshifted [O{\footnotesize III}] $\lambda$5007 emission lines. 
We also found that notable fractions of DRGs, HEROs, 24 $\mu$m sources, AGNs,  
and the counterparts of LABs and the AzTEC/ASTE 1.1-mm sources in the SSA22 field
are the members of the protocluster at $z=3.09$. 

The mass of the SSA22 protocluster was evaluated 
to be $2\sim5\times10^{14} ~M_{\odot}$.  
This structure is very likely to be one progenitor 
of the most massive clusters in the current Universe. 

We also confirmed the presence of massive quiescent galaxies 
in the SSA22 protocluster. 
It supports the early formation of massive early-type galaxies  
in such high density region. 
Most of them are also detected in X-ray, 
suggesting that AGNs play an important role in the 
formation history of massive early-type galaxies. \\ \\ 

This study is based on data collected at Subaru Telescope, 
which is operated by the National Astronomical Observatory of Japan. 
We would like to thank the Subaru Telescope staff 
for many help and support for the observations. 
Our studies owe a lot deal to the archival Subaru 
Suprime-Cam \citep{2004AJ....128..569M}, {\it Spitzer} IRAC \& MIPS data taken in
\citet{2009ApJ...692.1561W}, {\it Chandra} data taken in
\citet{2009ApJ...691..687L}. 
We also thank to AzTEC/ASTE observers of the SSA22 field 
providing the updated source catalog.  
This work was supported by Global COE Program "Weaving
Science Web beyond Particle-Matter Hierarchy", MEXT, Japan. 
YM acknowledges support from JSPS KAKENHI Grant Number 20647268. 
This work was partially supported by JSPS Grants-in-Aid for Scientific  Research No.26400217.

Funding for the SDSS and SDSS-II has been provided by the Alfred P. Sloan Foundation, the Participating Institutions, the National Science Foundation, the U.S. Department of Energy, the National Aeronautics and Space Administration, the Japanese Monbukagakusho, the Max Planck Society, and the Higher Education Funding Council for England. The SDSS Web Site is http://www.sdss.org/. The SDSS is managed by the Astrophysical Research Consortium for the Participating Institutions. The Participating Institutions are the American Museum of Natural History, Astrophysical Institute Potsdam, University of Basel, University of Cambridge, Case Western Reserve University, University of Chicago, Drexel University, Fermilab, the Institute for Advanced Study, the Japan Participation Group, Johns Hopkins University, the Joint Institute for Nuclear Astrophysics, the Kavli Institute for Particle Astrophysics and Cosmology, the Korean Scientist Group, the Chinese Academy of Sciences, Los Alamos National Laboratory, the Max-Planck-Institute for Astronomy, the Max-Planck-Institute for Astrophysics, New Mexico State University, Ohio State University, University of Pittsburgh, University of Portsmouth, Princeton University, the United States Naval Observatory, and the University of Washington.

\bibliographystyle{apj}
\bibliography{apj-jour,ssa22-moircs-spec}

\clearpage
\LongTables
\begin{landscape}
\begin{table*}[htbp]
\caption{        {\it Catalog of the galaxies confirmed by our NIR spectroscopic observations } 	}
 \begin{center}
\begin{tabular}{lccccccccc}
\\ \hline \hline
ID  & grism & $K_s$\footnotemark[a] & $J-K$\footnotemark[b] & $z_{\rm lit}$\footnotemark[c] & Detected Lines & $z_{\rm spec}$ & Flux\footnotemark[d]  & $M_{\rm star}$ & note \\
     &   & (mag) & (mag) &  & & & ($10^{-17}$ ergs s$^{-1}$ cm$^{-2}$)   & (10$^{10}~M_{\odot}$) &  
\\ \hline
J221718.0+001735.6 & VPH-$K$ & 23.5 & 1.1 & … & [O{\scriptsize III}] 5007 & 3.1423 $\pm$ 0.0003 & 12.4 $\pm$ 0.6 & $ 2.8_{- 0.5 }^{+ 1.2 } $ & ...\\
J221719.4+001657.1 & VPH-$K$ & 23.2 & 1.4 & … & [O{\scriptsize III}] 4959, 5007 & 3.3082 $\pm$ 0.0001 & 5.8 $\pm$ 0.3 & $ 0.8_{- 0.1 }^{+ 2.2 } $ & ... \\
J221716.4+001718.6 & VPH-$K$ & 23.1 & 0.8 & … & [O{\scriptsize III}] 5007 & 3.3552 $\pm$ 0.0001 & 5.8 $\pm$ 0.4 & $2.5 _{-0.4}^{+1.0}$  & ... \\
J221715.7+001906.2 & VPH-$K$ & $>25.2$ & ... & 3.1015 & [O{\scriptsize III}] 5007 & 3.1008 $\pm$ 0.0001 & 2.0 $\pm$ 0.3 & $0.2_{0.1}^{+0.3}$  & Y12\\
J221726.1+001232.3 & VPH-$K$ & 22.1 & 1.3 & … & H$\beta$, [O{\scriptsize III}] 4959, 5007  & 3.1000 $\pm$ 0.0003 & 13.8 $\pm$ 1.4 & $ 9.5_{- 2.5 }^{+ 7.5 } $ & LAB01\footnotemark[e]   \\
J221725.7+001238.7 & VPH-$K$ & 23.6 & 0.3 & ... & [O{\scriptsize III}] 4959, 5007 & 3.1007 $\pm$ 0.0002; & 2.0 $\pm$ 0.3 & $ 1.1_{- 0.5 }^{+ 0.6 } $ & LAB01\footnotemark[e]  \\
J221724.9+001117.5 & VPH-$K$ & 23.1 & 0.8 & ... & [O{\scriptsize III}] 5007 & 3.0689 $\pm$ 0.0002 & 4.2 $\pm$ 0.4 & $ 4.1_{- 2.0 }^{+ 3.0 } $ & LAB16\footnotemark[e]  \\
J221732.5+001131.2 & VPH-$K$ & 23.4 & 0.4 & 3.0767 (3.0677) & [O{\scriptsize III}] 5007 & 3.0680 $\pm$ 0.0003 & 1.7 $\pm$ 0.3\footnotemark[f] & $4.4_{-0.3}^{+0.2} $	& LAB30\footnotemark[e], S06 (S06) \\
J221732.5+001132.8 & VPH-$K$ & 23.0 & 0.9 & 3.0716 (3.0649)  & [O{\scriptsize III}] 5007 & 3.0687 $\pm$ 0.0004 & 0.7 $\pm$ 0.2\footnotemark[f] & $4.2_{-0.6}^{+0.2} $	& LAB30\footnotemark[e], S06 (S06)	\\
J221733.9+001106.5 & VPH-$K$ & 22.5 & 1.0 & ... & [O{\scriptsize III}] 5007 & 2.8596 $\pm$ 0.0001 & 3.9 $\pm$ 0.4 & $ 4.5_{- 1.1 }^{+ 5.3 } $ & ... \\
J221730.2+001120.7 & VPH-$K$ & 20.9 & 0.8 & ... & [O{\scriptsize III}] 5007 & 3.0667 $\pm$ 0.0002 & 3.8 $\pm$ 0.3 & $ 19.4_{- 0.0 }^{+ 0.0 } $ & ... \\
J221720.8+001831.0 & $HK500$ & 20.4 & 1.4 & 2.840 & [O{\scriptsize III}] 5007 & 2.9054 $\pm$ 0.0007 & 31.9 $\pm$ 1.6 & $ 95.1_{- 18.9 }^{+ 0.0 } $ & L09 \\
J221726.3+001957.2 & $HK500$ & 23.2 & 1.8 & ... & [O{\scriptsize III}] 5007 & 3.0431 $\pm$ 0.0006 & 11.5 $\pm$ 1.2 & $ 3.0_{- 1.2 }^{+ 3.1 } $ & ... \\
J221724.8+001803.7 & $HK500$ & 22.3 & 2.5 & ... & [O{\scriptsize III}] 5007 & 3.3868 $\pm$ 0.0010\footnotemark[g] & 3.8 $\pm$ 1.9 & $ 28.4_{- 9.2 }^{+ 35.0 } $ & ... \\
J221727.0+001746.4 & $HK500$ & 23.2 & 0.9 & ... & [O{\scriptsize III}] 5007 & 3.3235 $\pm$ 0.0004 & 4.9 $\pm$ 0.5 & $ 1.8_{- 0.9 }^{+ 0.4 } $ & ... \\
J221728.5+001807.4 & $HK500$ & 23.0 & 0.9 & ... & [O{\scriptsize II}] 3726 & 3.4281 $\pm$ 0.0012 & 9.1 $\pm$ 2.2 & $ 6.3_{- 1.4 }^{+ 1.1 } $ & ...\\
J221727.3+001809.5  & $HK500$ & 22.7 & 1.2 & 3.091(3.0855) &  H$\beta$, [O{\scriptsize III}] 4959, 5007 & 3.0861 $\pm$ 0.0009 & 10.3 $\pm$ 2.3 & $ 2.0_{- 0.8 }^{+ 0.7 } $ & S03 (P01) \\
J221734.9+001911.8 & $HK500$ & 23.3 & 1.2 & ... & [O{\scriptsize III}] 5007 & 3.0609 $\pm$ 0.0010 & 6.8 $\pm$ 1.3 & $ 0.8_{- 0.3 }^{+ 1.4 } $ & ... \\
J221737.3+001823.2 & VPH-$K$ & 22.5 & 2.8 & ... & H$\beta$, [O{\scriptsize III}] 4959, 5007 & 3.0851 $\pm$ 0.0001 & 19.1 $\pm$ 0.7 & $ 8.0_{- 2.9 }^{+ 5.7 } $ & AzTEC14\footnotemark[h,i] \\
 & &  &  & ... & [O{\scriptsize III}] 4959, 5007 & 3.0926 $\pm$ 0.0003 & 6.5 $\pm$ 0.8 &  ...  & \\
J221736.8+001818.2 & VPH-$K$ & 23.1 & 1.6 & ... & [O{\scriptsize III}] 5007 & 3.0854 $\pm$ 0.0003 & 2.8 $\pm$ 0.4 & $ 9.3_{- 5.0 }^{+ 19.7 } $ & AzTEC14\footnotemark[h]   \\
J221737.3+001816.0 & $HK500$ & 21.6 & 2.8 & ... &  ...\footnotemark[j]  & $2.7, 3.00-3.15$ & ...   & $ 25.4_{- 5.8 }^{+ 7.3} $ & AzTEC14\footnotemark[h]   \\
J221737.0+001820.4 & VPH-$K$ & 22.8 & 0.4 & ... & P$\beta$ & 0.5763 $\pm$ 0.0002 & 3.6 $\pm$ 1.0 & $ 0.3_{- 0.0 }^{+ 0.0 } $ & AzTEC14\footnotemark[h]   \\
J221732.0+001655.5 & VPH-$K$ & 22.2 & 2.6 & ... & [O{\scriptsize III}] 5007 & 3.0909 $\pm$ 0.0004 & 2.8 $\pm$ 0.7 & $ 11.5_{- 4.9 }^{+ 23.9 } $ & LAB12\footnotemark[e]  \\
J221736.5+001622.6 & VPH-$K$ & 20.4 & 0.5 & 3.084 & [O{\scriptsize III}] 5007 & 3.0945 $\pm$ 0.0008; & 2.1 $\pm$ 3.5 & $ 11.8_{- 0.0 }^{+ 0.0 } $ & S03 \\
J221731.8+001606.3 & VPH-$K$ & 23.0 & 1.7 & ... & H$\beta$, [O{\scriptsize III}] 5007 & 3.0981 $\pm$ 0.0002 & 11.3 $\pm$ 0.8 & $ 3.0_{- 1.5 }^{+ 3.2 } $ & ... \\
J221737.1+001712.4 & VPH-$K$ & 23.5 & 1.1 & ... & [O{\scriptsize III}] 5007 & 3.0899 $\pm$ 0.0004 & 7.2 $\pm$ 0.9 & $ 3.6_{- 0.8 }^{+ 8.6 } $ & ...\\
J221732.2+001502.3 & VPH-$K$ & 22.8 & 1.8 & ... & [N$_{\rm II}$] 6584, H$\alpha$ & 2.3252 $\pm$ 0.0002 & 8.7 $\pm$ 1.7 & $ 12.5_{- 5.0 }^{+ 2.3 } $ & ...\\
J221737.3+001630.7 & VPH-$K$ & 22.0 & 2.4 & ... & [O{\scriptsize III}] 5007 & 3.0888 $\pm$ 0.0004 & 6.1 $\pm$ 0.8 & $ 8.1_{- 0.0 }^{+ 1.3 } $ & ... \\
J221727.8+001736.6 & $HK500$ & 23.7 & 1.8 & 3.0922 & [O{\scriptsize III}] 5007 & 3.0916 $\pm$ 0.0012 & 4.0 $\pm$ 2.2 & $ 3.9_{- 2.0 }^{+ 0.9 } $ & Y12 \\
J221729.6+001918.6 & $HK500$ & 23.6 & 1.8 & 3.102 & [O{\scriptsize III}] 4959\footnotemark[k] & 3.1021 $\pm$ 0.0024 & 3.2 $\pm$ 2.5 & $ 1.3_{- 0.3 }^{+ 2.1 } $ & Y12 \\
J221728.3+001954.4 & $HK500$ & 22.6 & 0.6 & ... & H$\beta$, [O{\scriptsize III}] 4959, 5007 & 3.1019 $\pm$ 0.0005 & 6.1 $\pm$ 1.6 & $ 4.1_{- 2.0 }^{+ 2.5 } $ & ... \\
J221728.5+001822.2 & $HK500$ & 23.0 & 0.5 & ... & [O{\scriptsize III}] 5007 & 3.3148 $\pm$ 0.0004 & 7.8 $\pm$ 0.6 & $ 2.2_{- 0.0 }^{+ 0.7 } $ & ... \\
J221729.7+001715.2 & $HK500$ & 22.7 & 0.5 & ... & [O{\scriptsize III}] 5007 & 3.1390 $\pm$ 0.0056 & 4.2 $\pm$ 3.9 & $ 4.5_{- 0.0 }^{+ 1.2 } $ & ... \\
J221725.4+001716.9 & VPH-$K$ & 21.7 & 2.4 & 3.120 & H$\beta$ & 3.0482 $\pm$ 0.0003 & 5.6 $\pm$ 0.6 & $ 11.2_{- 2.1 }^{+ 1.5 } $ & (LAB35)\footnotemark[e], L09 \\
J221725.2+001805.7 & VPH-$K$ & 22.7 & 2.2 & ... & [O{\scriptsize III}] 5007 & 3.0973 $\pm$ 0.0004 & 1.7 $\pm$ 0.4 & $ 17.2_{- 1.9 }^{+ 27.8 } $ & AzTEC99\footnotemark[h] \\
J221722.3+001640.1 & VPH-$K$ & 20.5 & 0.3 & 3.360 (3.353) & H$\beta$, [O{\scriptsize III}] 4959, 5007 & 3.3544 $\pm$ 0.0001 & 39.0 $\pm$ 1.4 & $ 6.5 _{- 0.0 }^{+ 0.0 } $ & S03 (S03)\\
\hline
\end{tabular} 
\end{center}
\end{table*}
\clearpage
\end{landscape}
\addtocounter{table}{-1}
\clearpage
\LongTables
\begin{landscape}
\begin{table*}[htbp]
\caption{        {\it Continued}  }
 \begin{center}
\begin{tabular}{lccccccccc}
\\ \hline \hline
ID  & grism & $K_s$\footnotemark[a] & $J-K$\footnotemark[b] & $z_{\rm lit}$\footnotemark[c] & Detected Lines & $z_{\rm spec}$ & Flux\footnotemark[d]  & $M_{\rm star}$ & note \\
     &   & (mag) & (mag) &  & & & ($10^{-17}$ ergs s$^{-1}$ cm$^{-2}$)   & (10$^{10}~M_{\odot}$) &  
\\ \hline
J221721.9+001755.5 & VPH-$K$ & 22.9 & 0.9 & ... & H$\beta$, [O{\scriptsize III}] 4959, 5007 & 3.4075 $\pm$ 0.0001 & 11.3 $\pm$ 0.4 & $ 1.9_{- 1.1 }^{+ 0.9 } $ &  ... \\
J221724.3+001945.1 & VPH-$K$ & 22.6 & 0.3 & ... & [O{\scriptsize III}] 5007 & 3.1211 $\pm$ 0.0003 & 2.4 $\pm$ 0.4 & $ 2.6_{- 0.1 }^{+ 0.5 } $ &  ... \\
J221724.6+001836.5 & VPH-$K$ & 21.9 & 0.6 & ... & [O{\scriptsize III}] 5007 & 3.0847 $\pm$ 0.0006 & 2.3 $\pm$ 0.5 & $ 7.5_{- 0.3 }^{+ 1.5 } $ &  ... \\
\hline
\footnotetext[a]{ Total magnitudes, MAG\_AUTO estimated by SExtractor \citep{1996AAS..117..393B} on the $K_s$-band images. }
\footnotetext[b]{ $J-K$ colors measured in 1$''$.1 apertures centered 
at the coordinates on the $K$-band images. }
\footnotetext[c]{ The spectroscopic redshifts in literatures. 
The references of the spectroscopic redshifts are shown in the notes as follows;  
\citet{2001ApJ...554..981P} (P01); \citet{2003ApJ...592..728S} (S03); \citet{2006ApJ...651..688S} (S06)
\citet{2009MNRAS.400..299L} (L09); \citet{2012AJ....143...79Y} (Y12). 
$z_{\rm lit}$ cited from S03, S06 and Y12 were identified with Ly$\alpha$ 
while those identified with absorption features are also written in the brackets of  
J221732.5+001131.2, J221732.5+001132.8 and J221722.3+001640.1. 
$z_{\rm lit}$ cited from L09 were measured with optical spectroscopic obsrvations. 
$z_{\rm lit}$ cited from P01 (in the bracket of J221727.3+001809.5) 
was identified with nebular emission lines obtained with NIR spectroscopy. }
\footnotetext[d]{ Flux values of the strongest emission lines. 
The flux values of [O{\scriptsize III}] $\lambda$5007 or H$\alpha$ 
are shown for the objects detected of more than one emission lines. }
\footnotetext[e]{ 
The counterparts of LABs at $z=3.09$ \citep{2004AJ....128..569M}. 
J221725.4+001716.9 is not located within but very close to LAB35. }
\footnotetext[f]{The flux values of J221732.5+001131.2 
and J221732.5+001132.8 may be underestimated. 
Since they were observed within the same slit and located very close, 
their signals were subtracted by each other in the sky subtraction. 
And also, to acquire the both objects, we put the slit slightly off the center of J221732.5+001132.8. 
Actually, the flux of J221732.5+001132.8 is smaller than that obtained by \citet{2014ApJ...795...33E}. }
\footnotetext[g]{ The emission line is not likely to oiginate in this bright object  
since it was detected at slightly below the position of the faint continuum emission 
which may be originated in this bright object. }
\footnotetext[h]{The counterparts of the AzTEC/ASTE 1.1-mm sources 
\citep{2014MNRAS.440.3462U}. }
\footnotetext[i]{Showing a line profile with double peaks. }
\footnotetext[j]{Confirmed from the Balmer and 4000\AA~breaks in its continuum spectrum (described in \S~2). }
\footnotetext[k]{ We interpreted this emission line as [O{\scriptsize III}] $\lambda$4959 
since the offset from $z_{\rm lit}$ become much larger 
if we interpret this line as [O{\scriptsize III}] $\lambda$5007. 
The wavelength where its [O{\scriptsize III}] $\lambda$5007 is expected from the $z_{\rm lit}$ is buried by crowded OH airglow lines.  }
\end{tabular} 
     \label{tab:tableobject}
\end{center}
\end{table*}
\clearpage
\end{landscape}

\end{document}